# Resonant Propagation of Entangled Rhodium Mössbauer Gammas

Yao Cheng, Zhongming Wang
Department of Engineering Physics, Tsinghua University, Beijing 100084
**Email: yao@tsinghua.edu.cn

We report the resonant propagation of the long-lived Mössbauer gamma in the time-resolved Mössbauer spectroscopy. Recently, three entangled gammas emitted from the E3 rhodium Mössbauer transition has been proposed to interpret the extraordinary observations in the previous report. Further observation reported here is the dynamic beat of these entangled gammas at room temperature and 77K. Apparent beat anisotropy reveals their long-distance resonant propagation, which leads to suppressed Doppler shift of entangled photon transport in the Borrmann channel.
PACS: 33.45.+x, 76.80.+y, 03.67.Mn

We have reported a trapping state attributed to the gravitational redshift by assuming the valid nuclear resonant absorption [1]. Conventional wisdoms dispute the $10^{-19}$-eV Mössbauer resonance at room temperature due to the second order Doppler shift (hereinafter SODS) [2,3]. However, SODS is negligible between nuclei closely apart due to their collective vibration of the 100% isotopic abundance of $^{103}$Rh. Nuclear excitation distributes over a coherence length as nuclear exciton, where the nuclei share the gammas collectively [4]. As long as the resonant absorption occurs, we shall observe dynamic beat. When the Mössbauer transition rate is much higher than the nuclear Larmor frequency, isotropic emission and absorption are assumed. On the contrary, the long-lived Mössbauer transition rates such as $^{45}$Sc, $^{107}$Ag, $^{109}$Ag and $^{103}$Rh are slower than the Larmor frequency under the earth magnetic field. Angular distribution of Zeeman components corresponding to the magnetic field is no longer negligible. Conventional cross sections of nuclear resonant and photo-electric scatterings have the same order of magnitude for the 40-keV Mössbauer gamma in rhodium. No pronounced Mössbauer resonance of rhodium can be observed, unless the photo-electric cross section of nuclear Borrmann modes is reduced by orders of magnitude [5]. Hutton *et al.* [5] described the multi-beam nuclear Borrmann mode as a superposition of multiple passes. Our findings [6] have demonstrated that this particular Borrmann mode is consisted of three entangled gammas (hereinafter tri-γ) [7]. Here, we put forward the tri-γ formulation and its associated interpretation for the extraordinary observations against Doppler shift.

Assume that the entangled Lamb-Mössbauer factor $f_{LM}$ of tri-γ to be

$$f_{LM} \approx \left\langle \exp\left(i\mathbf{k}_1 \cdot \mathbf{r}\right) + \exp\left(i\mathbf{k}_2 \cdot \mathbf{r}\right) + \exp\left(i\mathbf{k}_3 \cdot \mathbf{r}\right) \right\rangle^2, \tag{1}$$

where $\mathbf{r}$ is the atomic displacement from its equilibrium position and (1) is averaged over the $\mathbf{r}$ ensemble. The entangled wavevector is $\mathbf{k} = \left(\mathbf{k}_1 + \mathbf{k}_2 + \mathbf{k}_3\right)/3$ in the particular Borrmann channel along the (1,1,1) direction of the rhodium fcc crystal to keep the rotational symmetry. Difference wavevector is a reciprocal lattice vector **G** to satisfy the Bragg condition of $\mathbf{k}_n - \mathbf{k}_m = \mathbf{G}$. Three gammas are grouped with particular polarization noted by |3σ⟩3 as illustrated in Fig. 1 and Fig. 15 of [8] to provide the cancellation of the entangled electric field $\mathbf{E}_{|3\sigma\rangle_3}$ at the lattice sites [5,8]. Consequently, the photo-electric attenuation of this |3σ⟩3 Borrmann mode is suppressed, but its nuclear resonant scattering is enhanced for the E3 transition through the $\Delta J_z = \pm 3$ components. The electric field of |3σ⟩3 has anti-symmetric increment in the transverse vicinity in (A-2), where the rapid photo-electric effect is possible. The Borrmann transmission is thus a function of temperature. Even averaging over the fast vibration, the slow multipolar interaction $\mathbf{J}\cdot\mathbf{E}_{|3\sigma\rangle_3}$ does not vanish [8,9], where **J** is the nuclear E3 transition current density. The Borrmann channel has the waveguide analogue, where lattice guides the photon propagation via Bragg reflection. Figure 2 provides the postulated picture, which has smooth variation of collective motion along the channel. While the lattice nuclei of the same isotope move together, the collective Hannon's anomalous emission [8] has fixed energy flow at the local emitting Rh site and *vice versa* at the local absorbing Rh' site disregarding their orientations. The adiabatic invariant photon flux of the Lorentz transformation suggests the invariant photon energy, which leads to the suppressed Doppler shift in channel. Equation (B-2) reveals that averaged entangled B fields at Rh and Rh' are invariant despite their local velocities relative to their comoving frames. Equation (1) in the Borrmann channel with suppressed time dilation then becomes

$$f_{LM}(\mathbf{k}) \approx 9e^{-\left\langle (\mathbf{k}\cdot\mathbf{r})^2 \right\rangle}, \tag{2}$$

where the number of 9 is the enhancement factor [5,8] and **r** is relative to its comoving frame.

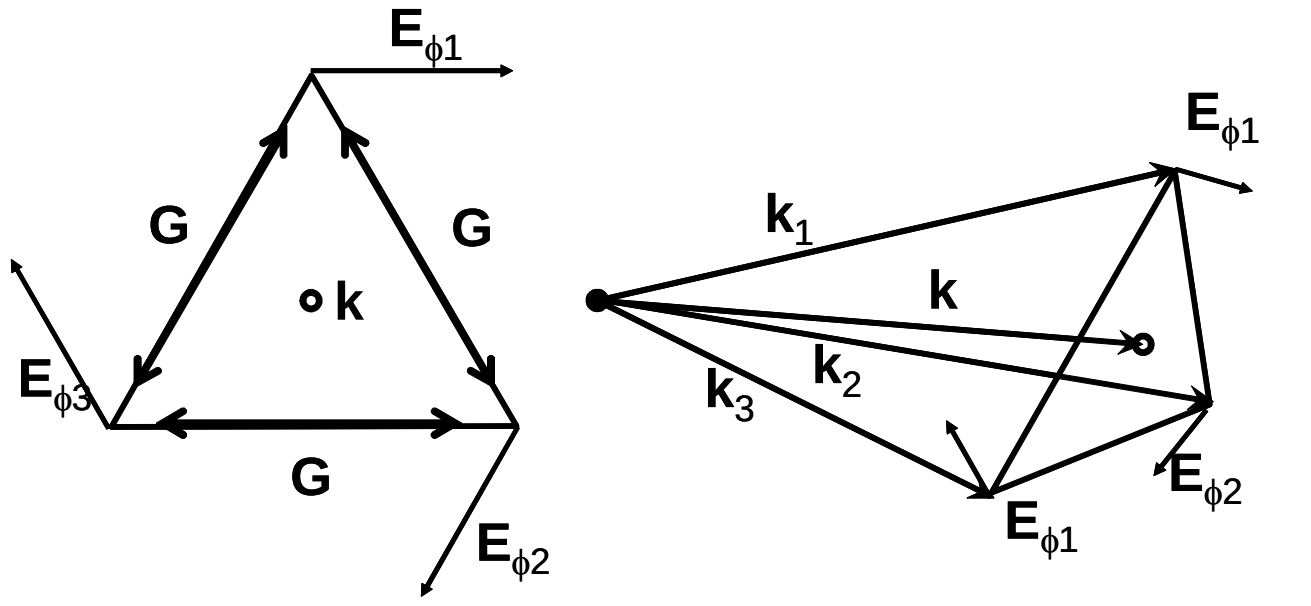

FIG. 1. The |3σ⟩3 Borrmann mode with propagating wave vector **k**. Difference wavevector **G** is the reciprocal lattice vector of the fcc rhodium crystal. The sum of electric field is cancelled at the lattice site (solid dot) of nodal plane [8].

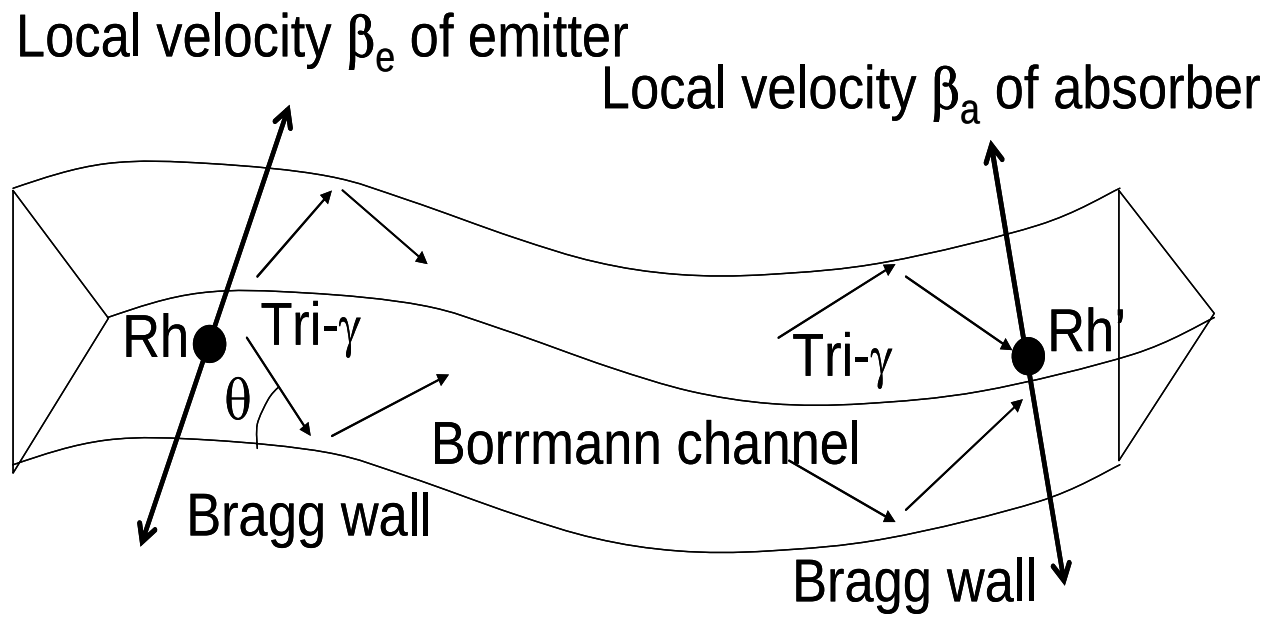

FIG. 2. Invariant photon energy flux in the Borrmann channel despite the velocity $\beta_e$ of emitter and $\beta_a$ of absorber. Tri-γ diffract at Bragg angle θ.

Propagation in a resonant medium is characterized by the decaying Lorentzian profile [4]. As a consequence, the time evolution of Mössbauer emission is no longer purely exponential [10], but containing an extra term $J_0^2\left(\sqrt{t/\tau_d}\right)$ of the Bessel function of the first kind of the order zero, where

$$\tau_d = \tau_0 / f_{LM}\mu_n\xi \tag{3}$$

is the beat time constant determined by the Lamb-Mössbauer factor $f_{LM}$, the extinction coefficient of nuclear resonant absorption $\mu_n$, and the lifetime $\tau_0$=4857s of the rhodium Mössbauer transition. We introduced here a scalar component ξ of the coherence length in this extremely sensitive Mössbauer effect. Periodic beat of the time-space argument of $\sqrt{t/\tau_d}$ is the unique feature of the resonant propagation. The double-hump energy distribution of the transmitted wave becomes narrower by increasing the number of independent scattering events, which leads to gradually enlarged time intervals of dynamic beat in (4) [4]. The fading interference between eroding humps is then demonstrated by the asymptotic Bessel oscillation

$$J_0\left(\sqrt{\frac{t}{\tau_d}}\right) \approx \sqrt{\frac{2\tau_d}{\pi t}} \cos\left(\sqrt{\frac{t}{\tau_d}} - \frac{\pi}{4}\right). \tag{4}$$

This oscillation can be observed using our pumping scheme as shown in (C-2). In our experiment, the sample size is 2.5 cm × 2.5 cm × 1 mm which is much larger than 50 μm and 22 μm corresponding to the non-Borrmann penetration depths for the photo-electric effect and the nuclear scattering respectively. Thus, $\xi$ in (3) should be restricted by the photo-electric penetration depth that leads to $\tau_d \sim \tau_0$ for $f_{LM} = 0.5$. In the Borrmann channel, $\tau_d$ is reduced by orders of magnitude due to enlarged parameters of $\xi$ and $f_{LM}$ of (2). The sample contains 1 mJ in the Mössbauer state after irradiation. The estimated initial decay of the thermal strain is $10^{-12}$/s using the rhodium properties listed in Table 1. The Doppler speed corresponding to one natural linewidth is 1fm/s. Thus, coherence length $\xi$ should be smaller than 1 mm and exponentially increasing in time, if Doppler shift were not suppressed in the Borrmann channel.

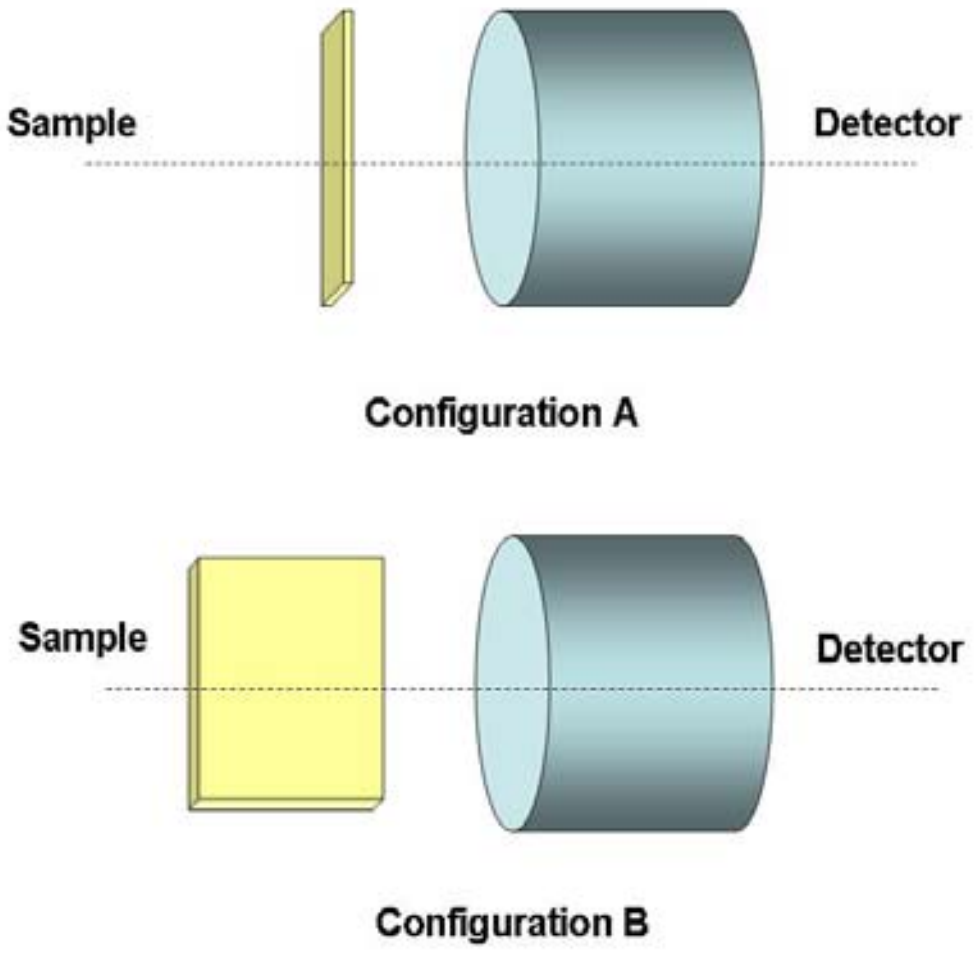

FIG. 3. Two configurations of the measurement set up. Configurations A and B have the 1-mm short axis and 2.5-cm long axis toward detector, respectively.

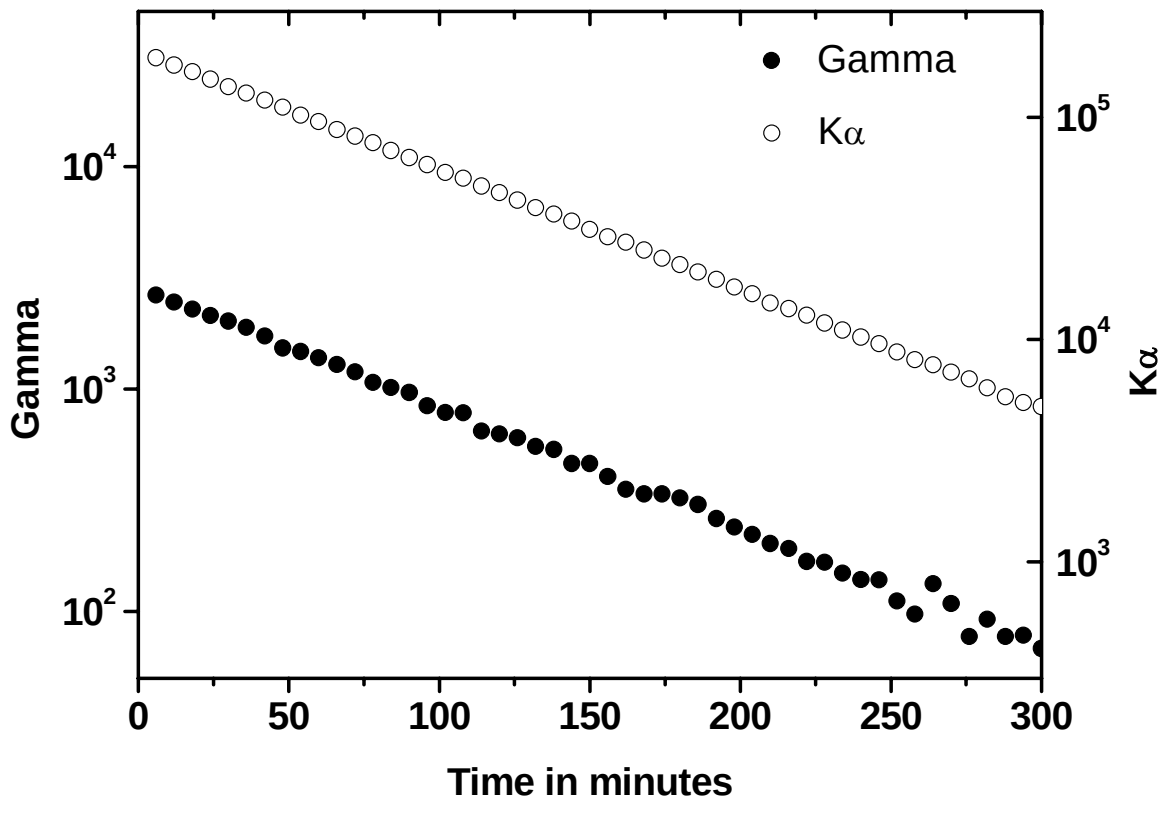

FIG. 4. Time evolution of rhodium emissions in the first case. Kα and Mössbauer γ are collected in 2-keV and 1-keV bandwidth, respectively. Each collection period is six minutes with the right ordinate for Kα and the left ordinate for γ.

The experimental conditions of configuration A such as detector and excitation procedure are detailed in [6]. The second and the third measurements have different sample preparation that two rhodium samples sandwiched by 0.5-mm tungsten and 0.05-mm plastic sheets are rotated by 90 degrees together. Tungsten sheets collimate the rhodium emission toward the detector. This configuration B is illustrated in Fig. 3. Three measurements are reported in this letter, i.e. the first case of uncooled configuration A, the second case of uncooled configuration B and the third case of cooled configuration B. In the third cooling case, we kept the sample in the room temperature of 18°C for twenty minutes and then poured in the liquid nitrogen to immerse sample from both tungsten sides from 20th minute to 213th minute. A typical time evolution of temperature has been reported previously in [6]. However, the quenching transient time becomes longer approximately to be several minutes due to additional heat capacity and thermal isolation.

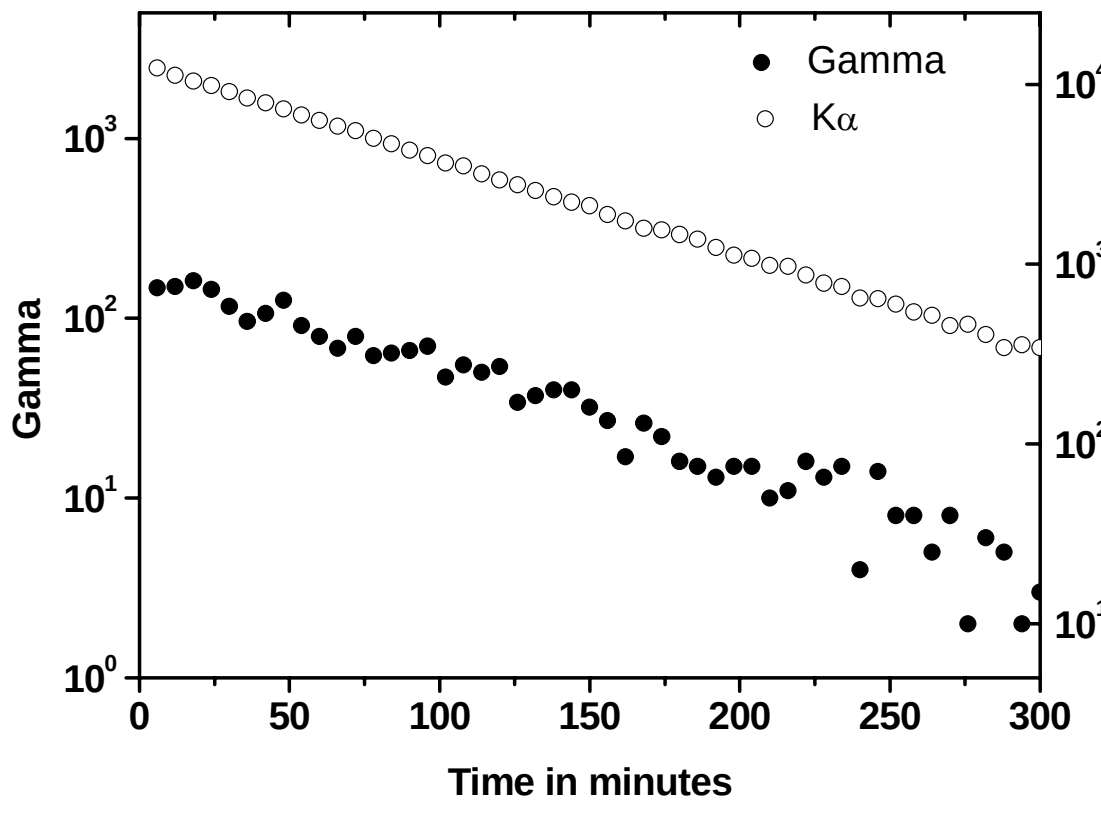

FIG. 5. Time evolution of rhodium emissions in the second case. Kα and Mössbauer γ are collected in 2-keV and 1-keV bandwidth, respectively. Each collection period is six minutes with the right ordinate for Kα and the left ordinate for γ.

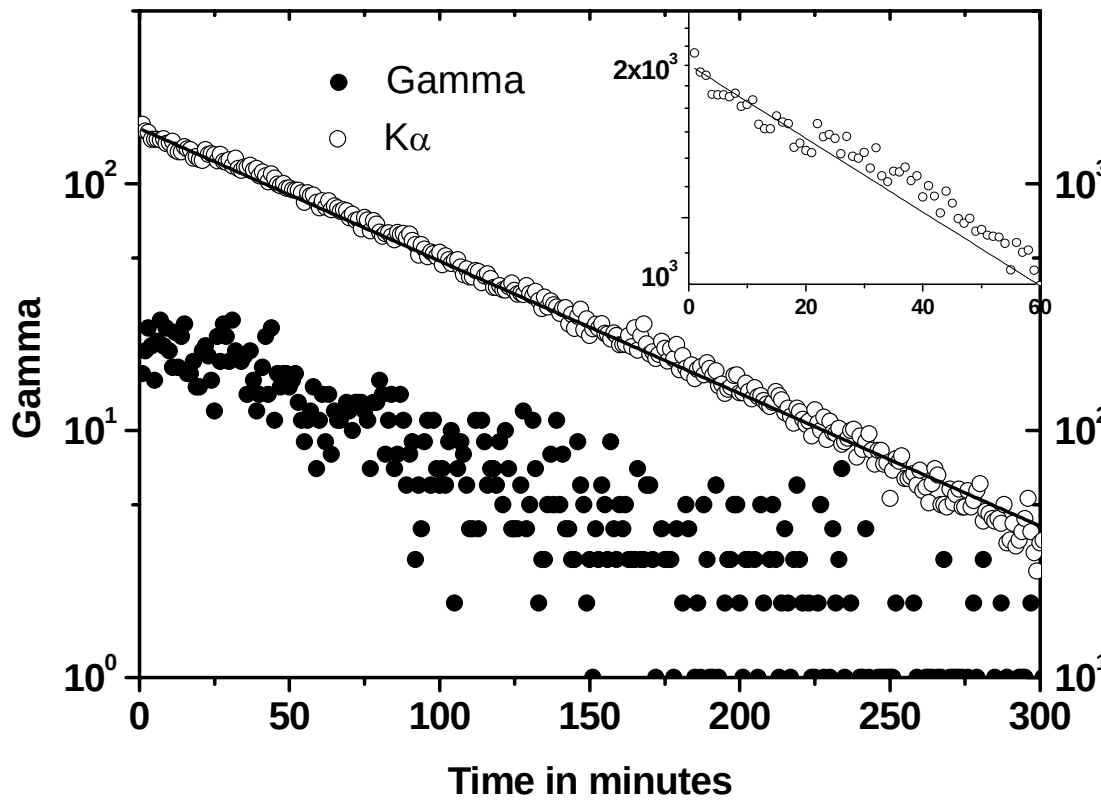

FIG. 6. Time evolution of rhodium emissions in the third case. Kα and Mössbauer γ are collected in 2-keV and 1-keV bandwidth, respectively. Each collection period is one minute with the right ordinate for Kα and the left ordinate for γ. The Kα counts have an enhanced count rate immediately after cooling as shown in the top right corner, and back to the normal count rate forty minutes after stop cooling.

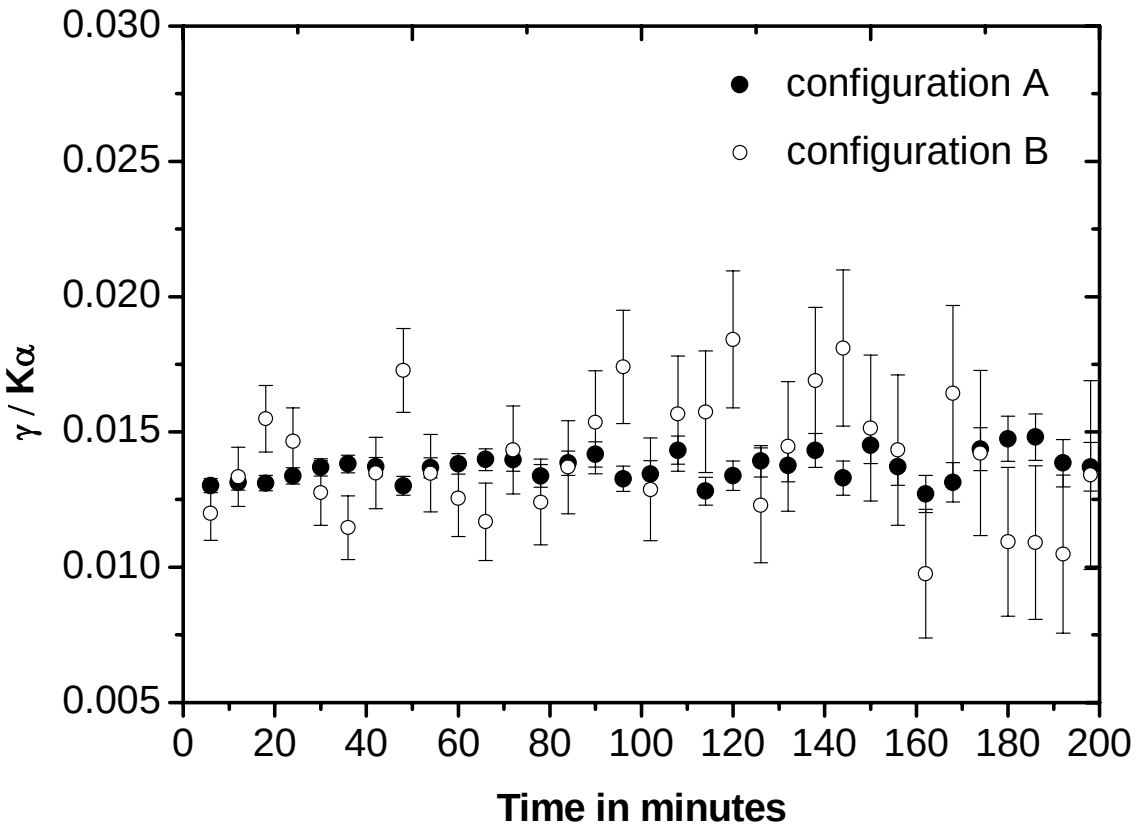

FIG. 7. Propagation beat of Mössbauer γ normalized by Kα counts measured in the first and second cases without cooling.

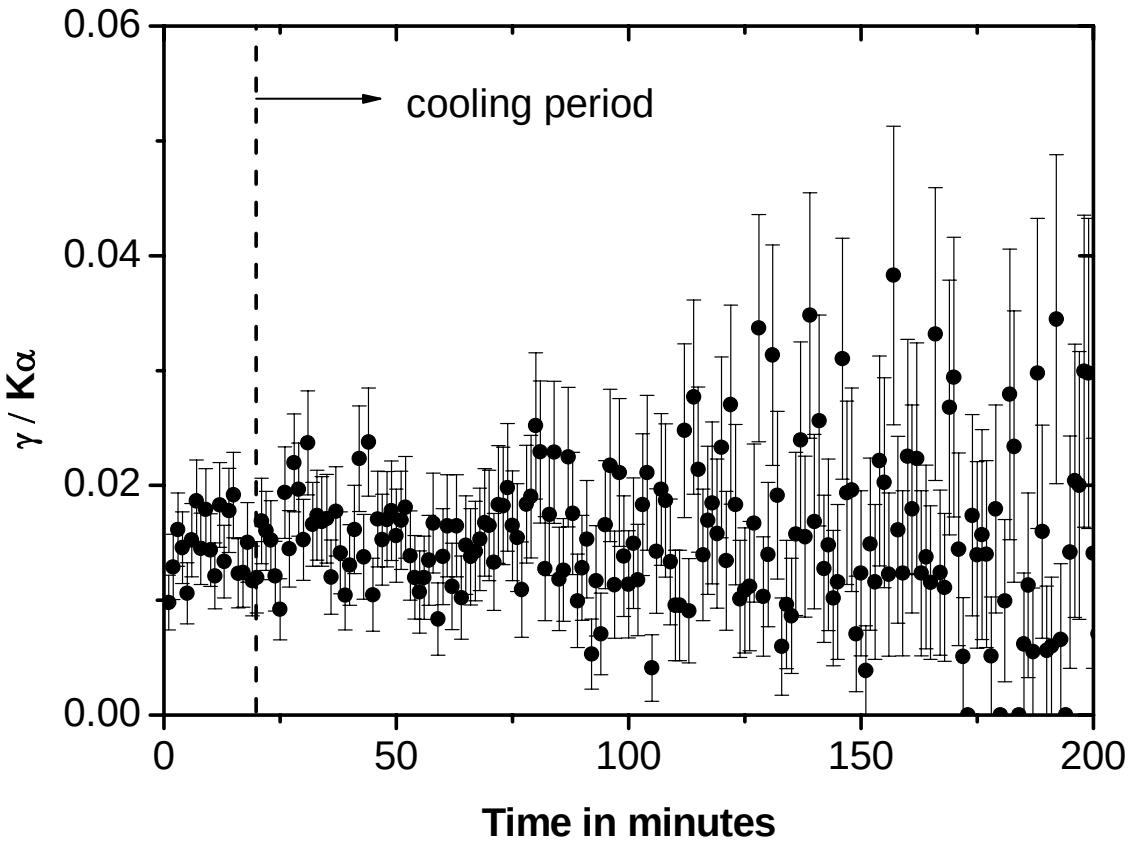

FIG. 8. Propagation beat of Mössbauer γ normalized by Kα counts of the third case with cooling.

| Expansion coefficient | Specific heat | Density |
|---|---|---|
| $8.5\times10^{-6}$ $K^{-1}$ | 244 $JK^{-1}$ $Kg^{-1}$ | 12.4 $gcm^{-3}$ |

TABLE 1. Rhodium properties at room temperature.

The time evolutions of the K X-rays and γ are illustrated in Figures 4, 5 and 6. Particular observation of interest is the slight increase of the K count after quenching in Fig. 6, which is blown up in the top-right corner. This increment reveals the inelastic scattering of the long-reaching tri-γ due to the ϕ-1.5 cm central irradiation spot. Thus, the Borrmann penetration depth under this cooling temperature shall be magnified to millimeters, which is 100-times larger than the non-Borrmann penetration depth. This increment of K counts in Fig. 6 returns to normal count rate later, roughly forty mutinies after quenching stop. The K count is suppressed in [6] instead of the enhancement reported here. These two different observations together reveal that the tri-γ superradiance enhanced from quenching is anisotropic, i.e. along the long sample axis. No coincidence count of gammas was ever measured up to now [6]. Tri-γ may be de-entangled before arriving detector.

We compare the time-resolved gamma counts of the two uncooled cases in Fig. 7. Gamma counts are normalized by $K\alpha_1$ counts due to the fact that gamma and $K\alpha_1$ have almost the same non-Borrmann photo-electric penetration depth (~50 μm) in rhodium. The apparent gamma beat disobeys (4). Obviously $\xi$ > 1mm, otherwise the same dynamic beat should be observed in Fig. 7. Figure 8 supports further this argument that $\xi$ > 1mm despite the strong quenching shrinkage. Gamma beat becomes more significant, when sample is cooled, even configured in A. Consequently, $\xi$ > 1 mm disproves the long-range SODS in the Borrmann channel. Furthermore, the first order Doppler shift is suppressed in the Borrmann channel too.

We thank Hong-Fei Wang for the manuscript preparation, Yanhua Shih, Yexi He for fruitful discussions, Bing Xia for data collection, Guozhen Wu and Mingda Li for the proofreading and Yuzheng Lin with his accelerator team.

**Appendix A**

Each entangled tri-gamma $|3\sigma\rangle_3$ is written in the cylindrical coordinate as

$$\mathbf{k}_n = k\sin\theta\mathbf{e}_\rho + k\cos\theta\mathbf{e}_z, \tag{A-1}$$

with three azimuth angles $\phi_n = (n-1)2\pi/3$, where n runs from 1 to 3. θ is the angle between the entangled wave vector $\mathbf{k} = \frac{1}{3}\sum_{n=1}^{3}\mathbf{k}_n = k\cos\theta\mathbf{e}_z$ and each single wavevector of (A-1). Field at the nuclear position **r**=(ρ,ϕ,z) is invariant under 2π/3 rotation as

$$\mathbf{E}_{|3\sigma\rangle_3}(\mathbf{r}) = Ee^{ikz\cos\theta}\sum_n \mathbf{e}_{\phi_n} e^{ik\rho\sin\theta\cos(\phi-\phi_n)}. \tag{A-2}$$

For simplicity take the z-axis of the nuclear spherical harmonic to coincide with the z-axis of the cylindrical coordinate of the tri-gamma. The interaction term [8] between this entangled field and the E3 nuclear transition at position (ρ,ϕ,z) is symmetric under 2π/3 rotation. Thus, the phase dependent part of the E3 octupole interaction is proportional to

$$\sum_n \left\langle \frac{7}{2}+,3,\pm3 \middle| (i\mathbf{k}_n\cdot\mathbf{r})^2\,\mathbf{J}(\mathbf{r})\cdot\mathbf{e}_{\phi_n} Ee^{i\mathbf{k}_n\cdot\mathbf{r}} \middle| \frac{1}{2}-,0,0 \right\rangle \propto \sum_n e^{i\mathbf{k}_n\cdot\mathbf{r}}, \tag{A-3}$$

where $\langle 7/2+,3,\pm3|$ is the excitation state of even parity, 7/2 spin, orbital angular momentum l=3, and magnetic quantum number m=±3 with threefold rotational symmetry. $|1/2-,0,0\rangle$ is the ground sate of odd parity with the corresponding quantum numbers. The angular dependency of nuclear orientation is omitted in the expression of (A-3). We obtain the Lamb-Mössbauer factor averaged over **r** uniformly distributed in ϕ without considering the nuclear orientation

$$f_{LM}(\mathbf{k}) = \left\langle \sum_n e^{i\mathbf{k}_n \cdot \mathbf{r}} \right\rangle^2 \approx 9e^{-\left\langle (\mathbf{k}\cdot\mathbf{r})^2 \right\rangle}. \tag{A-4}$$

**Appendix B**

The electromagnetic fields corresponding to two frames of the emitter Rh at rest and the moving absorber Rh' with velocity β (Fig. 2) can be written [9] as

$$\begin{aligned} \mathbf{E}' &= \gamma(\mathbf{E} + \boldsymbol{\beta} \times \mathbf{B}) - \frac{\gamma^2}{\gamma+1}\boldsymbol{\beta}(\boldsymbol{\beta}\cdot\mathbf{E}) \\ \mathbf{B}' &= \gamma(\mathbf{B} - \boldsymbol{\beta} \times \mathbf{E}) - \frac{\gamma^2}{\gamma+1}\boldsymbol{\beta}(\boldsymbol{\beta}\cdot\mathbf{B}) \end{aligned} \tag{B-1}$$

Here we apply the relativistic symbol of velocity $\boldsymbol{\beta}=\boldsymbol{\beta}_2-\boldsymbol{\beta}_1=\mathbf{v}_{21}/c$ and $\gamma = 1/\sqrt{1-\boldsymbol{\beta}\cdot\boldsymbol{\beta}}$. Two collective local comoving frames provide the conditions **E**=**E'**=0 and transverse $\mathbf{B}_\perp$=0 relative to the entangled wave vector **k** in (A-1) at lattice sites. Furthermore, in the vicinity of the lattice sites where electric field is available, the antisymmetric field distribution referred to the transverse direction in (A-2) vanishes by the **r** ensemble average. Vanishing E field and vanishing transverse B field of (B-1) lead to invariant longitudinal $\mathbf{B}_{||}$ field under Lorentz transformation

$$\boldsymbol{\beta}\cdot\langle\mathbf{B}'\rangle = \boldsymbol{\beta}\cdot\langle\mathbf{B}\rangle \quad \Rightarrow \quad \langle\mathbf{B}'\rangle = \langle\mathbf{B}\rangle. \tag{B-2}$$

**Appendix C**

Assume that the count rate of the resonant propagation has the form of

$$e^{-t/\tau_0}\cos^2\left(\sqrt{t/\tau_d} + \phi_0\right), \tag{C-1}$$

where $\tau_0$ and $\tau_d$ are defined in (4) and $\phi_0$ is an unknown phase. Sample is pumped by the bremsstrahlung from the time point of $-T_p$ to 0. We have the total count rate as incoherent accumulation

$$I(t) = N_0\int_t^{t+T_p} e^{-\tau/\tau_0}\cos^2\left(\sqrt{\tau/\tau_d} + \phi_0\right)d\tau, \tag{C-2}$$

where $N_0$ is the initial count rate corresponding to pumping efficiency.